\documentclass[reprint,aps,prl]{revtex4-2}

\usepackage{amsmath,amssymb,amsfonts,dsfont}
\usepackage{graphicx}
\usepackage[table]{xcolor}
\usepackage{hyperref}
\usepackage{subfigure}
\usepackage{dcolumn}
\usepackage{bm}
\usepackage{float}
\usepackage{mathtools}
\usepackage{amsthm}
\usepackage{bbm}
\usepackage{pxfonts}
\usepackage{pstricks}
\usepackage{verbatim}
\usepackage{booktabs}

\definecolor{orange}{cmyk}{0,0.5,1,0}
\definecolor{rossoCP3}{cmyk}{0,.88,.77,.40}
\definecolor{graa}{rgb}{0.8,0.8,0.8}
\definecolor{blaa}{rgb}{0.2,0.2,0.6}
\hypersetup{
	colorlinks, 
	bookmarksopen, 
	bookmarksnumbered,
	citecolor=blaa, 		
	linkcolor=rossoCP3,	
	urlcolor=rossoCP3,			
	    }

\newcommand{\beq}{\begin{eqnarray}}
\newcommand{\eeq}{\end{eqnarray}}

\newcommand{\softtheoryrule}{%
  \arrayrulecolor{black!28}%
  \specialrule{0.35pt}{2pt}{2pt}%
  \arrayrulecolor{black}%
}

\newcommand{\SU}{\mathrm{SU}}
\newcommand{\Sp}{\mathrm{Sp}}
\newcommand{\SO}{\mathrm{SO}}
\newcommand{\Spin}{\mathrm{Spin}}
\newcommand{\U}{\mathrm{U}}

\newcommand{\WWW}{W$^3$}

\newcommand{\plb}[1]{#1}

\begin{document}
\title{Perusing confining pseudoreal theories: \plb{hints} of emerging massless spin-1 bosons.}

\author{Giacomo Cacciapaglia}
\email{cacciapa@lpthe.jussieu.fr}
\affiliation{Laboratoire de Physique Theorique et Hautes Energies {\color{rossoCP3}LPTHE}, UMR 7589, Sorbonne Universit\'e \& CNRS, 4 place Jussieu, 75252 Paris Cedex 05, France.}
\author{Konstantinos Kollias}
\email{konstadinos.kollias@gmail.com}
\affiliation{Laboratoire de Physique Theorique et Hautes Energies {\color{rossoCP3}LPTHE}, UMR 7589, Sorbonne Universit\'e \& CNRS, 4 place Jussieu, 75252 Paris Cedex 05, France.}

\begin{abstract}
Solving quantum field theory, which is at the basis of the standard model of particle interactions, is one of the main tasks of contemporary theoretical physics. Minimal asymptotically free pseudoreal theories, containing one or two species of massless Weyl fermions in pseudoreal representations of the gauge group, flow towards a poorly understood infrared dynamics. We provide a comprehensive study of all \plb{non-supersymmetric} pseudoreal theories, \plb{first} identifying the ones that likely flow towards a conformal dynamics. \plb{Assuming that the remaining ones confine,} we \plb{find that tumbling hints the possible formation of} a non-trivial condensate dynamics\plb{, consistently with other symmetry considerations}. All but one one-species theories \plb{may} feature \plb{at least one} massless spin-1 state. Instead, the only confining two-species theory likely features two massless spin-1 states and one Nambu-Goldstone boson. These \plb{expectations must} be further tested, for instance by use of Lattice simulations, the functional renormalization group, and supersymmetric analogs. 
\end{abstract}

\maketitle

Local gauge symmetry in quantum field theory~\cite{Yang:1954ek} is the most advanced device leading our understanding of fundamental particles and their interactions. The dynamics of gauge interactions features a rich array of phenomena, among which spontaneous breaking via a scalar field vacuum~\cite{Englert:1964et,Higgs:1964pj} and asymptotic freedom~\cite{Politzer:1973fx,Gross:1973id} with confinement, both being realized in the standard model of particle physics. Asymptotically free gauge theories are particularly interesting as they are valid up to arbitrarily high energies and they can be easily constructed (as compared to asymptotically safe theories~\cite{Weinberg:1980gg}, which are rare~\cite{Litim:2014uca}). They are characterized by a gauge coupling that renormalizes to zero in the deep ultraviolet (UV), leading to a UV free theory. Conversely, the theory runs to strong coupling in the infrared (IR). What actually happens to such theories in the strongly coupled IR remains a mystery, notwithstanding decades of investigations and the realization in Nature within quantum chromo-dynamics (QCD): confinement can be modeled, but it has not been proven from first principles~\cite{Gross:2022hyw}. This fact undermines our full understanding of quantum field theory.

In this work, we focus on scalarless gauge theories with massless charged fermions: such theories can be classified based on the properties of the fermion representations under the gauge group.
Vector-like gauge theories feature Dirac fermions in complex representations or Weyl fermions in real representations, which admit gauge-invariant mass terms and are characterized by a real and positive Euclidean path integral. Their IR behavior can be delineated thanks to powerful exact results, most notably those of Vafa and Witten \cite{Vafa:1983tf,Vafa:1984xg}. They state that their vector-like global symmetries cannot be broken, consistently with the formation of a gauge-invariant bilinear fermion condensate. This may break the axial symmetries of the fermionic sector and lead to (light) massless (pseudo-)Nambu-Goldstone degrees of freedom: this is realized  in QCD via the light pions.
Instead, chiral gauge theories~\cite{Eichten:1985ft,Eichten:1985fs} feature Weyl fermions in complex representations within anomaly-free combinations that do not admit gauge-invariant mass terms, and are characterized by a complex path integral. Considerable effort has been devoted to understanding their IR dynamics, in particular for the classic examples of the Georgi-Glashow~\cite{Georgi:1974sy} and Bars-Yankielowicz~\cite{Bars:1981se} models, and their variants, see e.g.~\cite{Eichten:1985fs,Sannino:2008pz,Sannino:2009za,Poppitz:2009kz,Poppitz:2009uq,Bolognesi:2019wfq,Bolognesi:2020mpe,Bolognesi:2021hmg,Csaki:2021aqv,Csaki:2021xhi,Bolognesi:2021jzs,Li:2025tvu,Li:2026ayh}. In particular, the gauge symmetry may be spontaneously broken by fermion condensates~\cite{Raby:1979my}, in the so-called Higgsed phase. Anomaly matching {\`a} la 't~Hooft~\cite{tHooft:1976rip,tHooft:1976snw,tHooft:1979rat} also provides a useful tool to constrain the IR spectrum of chiral theories.
In both classes, depending on the fermion content of the theory, the gauge coupling may run towards an IR fixed point~\cite{Banks:1981nn} if the one-loop beta function is small, leading to a quantum conformal field theory: this phenomenon usually occurs for large fermion degeneracy~\cite{Dietrich:2006cm,Sannino:2008pz,Sannino:2009aw,Ryttov:2009yw,Mojaza:2012zd} or large representations~\cite{Dietrich:2006cm}.

\vspace{0.5cm}

A third class can be identified that lies in between, in some sense, being defined in terms of an odd number of Weyl fermions in a pseudoreal representation of the gauge group. While a fermion mass term is permitted by the gauge symmetry, one Weyl fermion always remains massless. Such \textit{pseudoreal theories} share with vector-like theories a real Euclidean path integral, while no non-trivial mass operator can be added, like in chiral theories.
The contradictory properties of pseudoreal theories were partly resolved by Witten~\cite{Witten:1982fp}, who showed that the prototypical example of a pseudoreal gauge theory (an $\SU(2)$ gauge theory with a single Weyl fermion in the defining representation)  is inconsistent due to a global topological anomaly. This phenomenon is known to extend to $\Sp(2N)$ gauge theories with fermions in pseudoreal representations of half-integer Dynkin index~\cite{Okubo:1989vn,OkuboZhang1989}. In addition, the 3-index symmetric representation of the symplectic family  is also known to exhibit a new type of anomaly -- namely the Wang-Wen-Witten (\WWW) anomaly \cite{Wang:2018qoy}. The \WWW anomaly, however, depends on some global assumptions on the theory which may not be met in general. No other anomaly can exist besides the Witten and \WWW anomalies~\cite{Wang:2018qoy}, thus leaving us with a handful of asymptotic free pseudoreal theories, which were first classified in~\cite{Cacciapaglia:2026kry}.

\vspace{0.5cm}

In this letter we aim at defining a simple consistent framework for understanding the possible IR dynamics of these pseudoreal theories~\cite{Cacciapaglia:2026kry}, summarized in Table~\ref{tab:pseudoreal_summary}. We consider the minimal configuration comprising a single Weyl fermion without Witten anomaly~\cite{Witten:1982fp}, while we keep theories affected by the \WWW anomaly~\cite{Wang:2018qoy}. Some theories based on $\Sp(2N)$ are defined in terms of two fermions with semi-integer Dynkin index to cancel the Witten anomaly~\cite{Cacciapaglia:2026kry}. All theories possess a single anomalous global $\U(1)_A$ symmetry, while only the two-species theories have an anomaly-free $\U(1)_V$ symmetry; also, due to the massless fermions, the topological $\theta$-angle can always be rotated away by $\U(1)_A$.
If the Vafa-Witten theorem covered such theories, it would simply tell us that parity cannot be broken in the IR, however it would remain an open question if any condensate forms and if the gauge symmetry breaks or not (recall that no gauge-invariant bi-fermion scalar can be constructed). Discrete and generalized symmetries may also play an important role in determining the IR fate of these theories~\cite{Csaki:1997aw,Aharony:2013hda,Gaiotto:2014kfa}. Only two theories of this class have been considered in the literature: $\SU(2)$ with a single 3-index symmetric ${\bf 4}$~\cite{Poppitz:2009kz} (\WWW-anomalous) and $\SU(6)$ with a 3-index antisymmetric ${\bf 20}$~\cite{Yamaguchi:2018xse,Bolognesi:2019fej,Bolognesi:2021hmg,Yamaoka:2025eej}.
In both cases, arguments based on discrete symmetries seem to point towards a confined theory where a condensate made of four~\cite{Yamaguchi:2018xse} or two~\cite{Bolognesi:2021hmg} fermions would be needed to break the discrete chiral symmetry.
Our analysis, in addition, points towards an IR dynamics characterized by a massless spin-1 free state in all cases, except one.

\vspace{0.5cm}

To substantiate this conclusion, we follow the methodology sketched below:
\begin{itemize}
    \item[\textit{i})] Determine which theories are likely to flow to a conformal IR fixed point, by employing an ansatz for the all-order beta function~\cite{Ryttov:2007cx,Pica:2010mt} \plb{further confirmed via an independent three-loop analysis}. In this case, the IR dynamics is described in terms of a conformal field theory, populated by gauge-invariant operators.
    \item[\textit{ii})] Analyze the remaining theories via the tumbling approach \cite{Raby:1979my,Dimopoulos:1980hn}, which postulates the formation of a gauge-breaking bilinear. Note that this description may be equivalent to that of a gauge-invariant four-fermion condensate, so that the gauge symmetry is only apparently broken. This phenomenon is similar to spontaneous breaking due to the Higgs mechanism.
    \item[\textit{iii})] Study the discrete and 1-form generalized global symmetries of the theories, analogously to~\cite{Yamaguchi:2018xse,Bolognesi:2019fej,Bolognesi:2021hmg}.
    \item[\textit{iv})] Analyze the structure of the lowest dimension gauge-invariant operators and identify candidates for the massless state.
    \item[\textit{v})] Match the results of steps \textit{ii}, \textit{iii} and \textit{iv} to obtain a consistent picture.
\end{itemize}
As we will see, all pseudoreal theories outside the conformal window share the following features: the scalar bilinear always contains the adjoint representation of the gauge group; the lowest dimensional gauge invariant operator is a spin-1 fermion current; no composite fermionic operator exists, hence all anomalous global symmetries must be broken \plb{unless they can be matched by a topological quantum field theory (TQFT)}; the discrete symmetries require the existence of fermion condensates when global anomalies are present.

\begin{table*}
\centering
\begin{tabular}{@{}llccllll@{}}
\toprule
\textbf{$\mathcal{G}$} &
\textbf{Matter} &
\textbf{SUSY?} &
\textbf{CW} &
\textbf{Tumbling subgroup} &
\textbf{$R$} &
\textbf{$\Delta$} &
\textbf{Massless content / annotations}
\\
\midrule

$\SU(2)\sim\Sp(2)$ &
$\mathbf{4}_{S3}$ (*) &
No &
likely &
-- &
-- &
-- &
--
\\

$\Sp(4)$ &
$\mathbf{16}_{R11}$ &
Yes &
? &
$\SU(2)\times \U(1)$ &
$\mathbf{10}$ &
$9/2$ &
$\U(1)$
\\

$\Sp(6)$ &
$\mathbf{56}_{S3}$ (*) &
No &
likely &
-- &
-- &
-- &
--
\\

$\Sp(6)$ &
$\mathbf{64}_{R11}$ &
No &
likely &
-- &
-- &
-- &
--
\\

$\Sp(8)$ &
$\mathbf{48}_{A3}$ &
Yes &
? &
$\SU(4)\times \U(1)$ &
$\mathbf{36}$ &
$23/2$ &
$\U(1)$
\\

$\Sp(10)$ &
$\mathbf{132}_{A5}$ &
No &
likely &
-- &
-- &
-- &
--
\\

$\Sp(12)$ &
$\mathbf{208}_{A3}$ &
Yes &
likely &
-- &
-- &
-- &
--
\\

$\Sp(16)$ &
$\mathbf{544}_{A3}$ &
No &
likely &
-- &
-- &
-- &
--
\\

\midrule

$E_7$ &
$\mathbf{56}_{F}$ &
Yes &
out &
$E_6\times \U(1)$ &
$\mathbf{133}$ &
$21/2$ &
$\U(1)$
\\

\midrule

$\SU(6)$ &
$\mathbf{20}_{A3}$ &
Yes &
out &
$\SU(5)\times \U(1)$ &
$\mathbf{35}$ &
$9/2$ &
$\U(1)$
\\
&
&
&
&
$\SU(4)\times \SU(2)\times \U(1)$ &
$\mathbf{35}$ &
$9/2$ &
$\U(1)$ +  $(\mathbf{6},\mathbf{2})_0$; tumbles
\\
 &
 &
 &
 &
$\SU(3)\times \SU(3)\times \U(1)$ &
$\mathbf{35}$ &
$9/2$ &
$\U(1)$\\
\midrule

$\Spin(11)$ &
$\mathbf{32}_{Spin}$ &
Yes &
out &
$\Spin(10)$ &
$\mathbf{11}$ &
$35/4$ &
None
\\
&
&
&
&
$\Spin(9)\times \U(1)$ &
$\mathbf{55}$ &
$19/4$ &
$\U(1)$; N-MAC
\\

\softtheoryrule

$\Spin(12)$ &
$\mathbf{32}_{Spin}$ &
Yes &
out &
$\SU(6)\times \U(1)$ &
$\mathbf{66}$ &
$13/2$ &
$\U(1)$ +  $\mathbf{20}_0$; tumbles
\\
&
&
&
&
$\Spin(10)\times \U(1)$ &
$\mathbf{66}$ &
$13/2$ &
$\U(1)$
\\

\softtheoryrule

$\Spin(13)$ &
$\mathbf{64}_{Spin}$ &
Yes &
out &
$\Spin(11)\times \U(1)$ &
$\mathbf{78}$ &
$17/2$ &
$\U(1)$
\\
&
&
&
&
$\Spin(10)\times \SU(2)$ &
$\mathbf{286}$ &
$9/2$ &
None; N-MAC
\\

\softtheoryrule

$\Spin(19)$ &
$\mathbf{512}_{Spin}$ &
No &
likely &
-- &
-- &
-- &
--
\\

$\Spin(20)$ &
$\mathbf{512}_{Spin}$ &
No &
likely &
-- &
-- &
-- &
--
\\

\midrule

$\Sp(4)$ &
$\mathbf{4}_F+\mathbf{20}_{S3}$ (*) &
No &
likely &
-- &
-- &
-- &
--
\\

\softtheoryrule

$\Sp(6)$ &
$\mathbf{6}_F+\mathbf{14}'_{A3}$ &
Yes &
out &
$\SU(3)\times \U(1)$ &
$\mathbf{21}$ &
$7/2$ &
NGB + $\U(1)$
\\
&
&
&
&
$\SU(2)\times \Sp(4)$ &
$\mathbf{14}$ &
$5/2$ &
NGB; tumbles again N-MAC
\\
&
&
&
&
$\SU(2)\times \U(1)\times \U(1)$ &
 $\bf 21$ + $\bf 14$ &
 &
NGB + U(1) + U(1); strong anomaly
\\

\softtheoryrule

$\Sp(10)$ &
$\mathbf{10}_F+\mathbf{110}_{A3}$ &
Yes &
likely &
-- &
-- &
-- &
--
\\

$\Sp(14)$ &
$\mathbf{14}_F+\mathbf{350}_{A3}$ &
No &
likely &
-- &
-- &
-- &
--
\\

\bottomrule
\end{tabular}
\caption{Summary of pseudoreal gauge theories~\cite{Cacciapaglia:2026kry} defined in terms of the gauge group $\mathcal{G}$ and the representations of the one or two fermion species, where the (*) labels theories affected by the \WWW anomaly~\cite{Wang:2018qoy}. We indicate if the theory has a supersymmetric asymptotically free counterpart (third column); conformal-window estimates (fourth); and properties under tumbling (fifth to 8th columns). Tumbling data are displayed only for theories that are likely non-conformal (out) or whose conformal status is uncertain (?). In the last column, $\U(1)$ indicates the presence of a massless spin-1 state, NGB that of a Nambu-Goldstone boson from the breaking of the anomaly-free $\U(1)_V$. We recall that $\Spin(N)$ is the double cover of $\SO(N)$ containing the spinorial representation.}
\label{tab:pseudoreal_summary}
\end{table*}

\vspace{0.5cm}

\subsubsection*{Sitting below the conformal window}
The existence of IR fixed points for gauge-fermion theories was first proven in the weak coupling regime~\cite{Banks:1981nn}. Typically, the IR conformal behavior starts when the number of fermions is right below the loss of asymptotic freedom, and ends when a minimal multiplicity is reached. It is generally hard to determine the lower edge of the conformal window~\cite{Dietrich:2006cm} in terms of the fermion degeneracy as this typically occurs at strong coupling.
Dealing with non-supersymmetric theories, we need to rely on some approximation: here we follow an ansatz for the beta function, based on the supersymmetric exact result \cite{Novikov:1983uc}, first proposed in \cite{Ryttov:2007cx}. The conjectured all-orders beta function can be expressed in the form \cite{Pica:2010mt}:
\begin{equation}
    \beta(g) = - \frac{g^3}{16 \pi^2} \frac{\beta_0 +\frac{1}{3} \sum_r  n_r T_r\ \gamma_r(g^2)\ \Delta_r}{1-\frac{g^2}{8\pi^2} \frac{17}{11} C_G}\,,
\end{equation}
where
\begin{equation}\label{eq:b}
\beta_0 = \frac{11}{3} C_G - \frac{2}{3} \sum_r n_r T_r\,, \qquad \Delta_r = 1 + \frac{7}{11} \frac{C_G}{C_r}\,,
\end{equation}
and $r$ is the Weyl fermion representation under the gauge group ($T_r$ and $C_r$ are the Dynkin index and Casimir of $r$, respectively, while $C_G$ is the Casimir of the adjoint). The anomalous dimension $\gamma_r^\ast$ at the putative IR fixed point is related to the dimension $D$ of the bilinear fermion mass operator as $D(\psi_r^2) = 3 + \gamma_r^\ast > 1$, hence it is bound to be $\gamma_r^\ast > -2$.
Violation of the above inequality for at least one species $\psi_r$ would signal that the IR fixed point does not exist. Assuming the existence of the fixed point, $\gamma_r^\ast$ can be computed from $\beta(g^\ast)=0$ for one fermion species so that the bound $\gamma_r^\ast > -2$ implies
\begin{equation}
    n_r>\frac{121C_GC_r}{2T_r(22C_r+7C_G)}\,.
\end{equation}
If the inequality above is violated, then the fixed point assumption is inconsistent and the theory is outside the conformal window; else, it's likely inside. Note, however, that the bound cannot be relied upon for $n_r=1$, as the gauge-invariant $\psi_r\psi_r$ operator vanishes and $\gamma_r^\ast$ is ill-defined. 

Henceforth, we can distinguish the following cases:
\begin{itemize}
    \item  If the inequality is violated for $n_r = 2$, then $n_r=1$ must also be \textit{out}side the conformal window.
    \item  If the inequality is valid even for $n_r=1$, then the theory is \textit{likely} to be conformal.
    \item In the intermediate case, where $n_r=1$ violates the inequality but $n_r=2$ respects it, it's not possible to give any confident indication.
\end{itemize}
For the third point, in principle one could study the non-vanishing $\psi_r^4$ operator, however in this case there is no exact relation between $D(\psi_r^4)$ and $\gamma^\ast_r$. The aforementioned analysis also extends to two-species theories, where it is enough that one species violates the inequality $\gamma^*_{r_i}>-2$ for the theory to be outside the conformal window. Using these, we can determine which of the pseudoreal gauge theories are \plb{most likely} non-conformal in the IR. The results are summarized in the fourth column of Table \ref{tab:pseudoreal_summary}, where a ``?'' tags theories that cannot be determined. Interestingly, all theories affected by the \WWW anomaly are likely to give an IR conformal dynamics.

\plb{As an independent perturbative check, we also computed the 2-loop and 3-loop beta function coefficients of the pseudoreal theories. The former is scheme independent, while the latter is taken in the $\overline{\text{MS}}$-scheme. We find that the beta function of all eleven theories labeled as ``likely'' conformal has a positive infrared-attractive zero at both orders. On the other hand, all 6 theories labeled ``out'' do not exhibit such zeroes. The 2 theories for which the preceding criterion is inconclusive behave differently: $\Sp(4)$ with a $\mathbf{16}$ has zeros at both orders but their values differ substantially, while $\Sp(8)$ with a $\mathbf{48}$ has no zero at both order. The coefficients and candidate fixed point couplings are summarized in the Appendix. }

\subsubsection*{Tumbling analysis}
Having sifted through the list of pseudoreal theories, we now proceed with the tumbling analysis for those likely yielding a non-conformal IR dynamics, or that remain undetermined. The main idea is to identify all the spin-0 bilinear fermionic operators, regardless of their gauge quantum numbers~\cite{Raby:1979my,Dimopoulos:1980hn}, and then establish which one is most likely to generate a condensate, i.e. the most attractive channel (MAC). For vector-like theories, this method naturally points towards the gauge singlet, which breaks the global symmetries respectfully to the Vafa-Witten theorem.

For pseudoreal theories, since no scalar gauge invariant bilinear operators can be constructed from $\psi_r$ alone, the lowest dimensional gauge invariant scalar operators are made of 4 fermions. A basis can be constructed as:
    \begin{equation}\label{eq:XR}
        X_R = (\psi_r \psi_r)_R (\psi_r \psi_r)_R\,,
    \end{equation}
where $(\psi_r \psi_r)_R$ indicates a scalar operator in the representation $R = \text{Sym} (r\otimes r)$ of the gauge group, deriving from the symmetric combination of $r$. Hence, the dynamical scale that gaps the theories in the IR can be associated to a condensate of one of the operators $X_R$. Following the tumbling analysis \cite{Dimopoulos:1980hn}, one could then formally assign a vacuum expectation value (VEV) to the scalar components $(\psi_r \psi_r)_R$, which then breaks the gauge group in a tumbling manner, until vector-like confinement occurs.\\

Due to the pseudoreal nature of $r$, all models contain a common channel $R=\text{Adj}$. This is also the smallest $R$ and therefore, as we will see, always the MAC (except in one case) for the one-species theories. Note that this is never the case for vector-like and chiral theories: in the former case, the singlet is always present and more attractive, while in the latter the adjoint never appears due to the chiral nature of the fermion representations. Now, assigning a VEV to the adjoint channel always breaks the gauge group $\mathcal{G}$ to
\begin{equation}
    \mathcal{G} \to \mathcal{H} \times \U(1)_X\,,
\end{equation}
where at least one abelian factor $\U(1)_X$ is always present, associated to the generator aligned with the VEV. Conversely, $\mathcal{H}$ is a group made of the remaining generators that commute with the VEV.
Typically, the fermion $\psi_r$ decomposes in a vector-like representation of $\mathcal{H}$, which then confines without massless states. Instead, the abelian gauge boson associated with $\U(1)_X$ remains massless. Hence, this simple analysis points towards an IR dynamics made of a \textit{non-interacting massless spin-1 state}.

More precisely, the number of unbroken abelian $\U(1)$ factors depends on the structure of the adjoint VEV, with the possibility of full abelianization~\cite{Bolognesi:2017pek,Bolognesi:2019wfq} leading to a number of massless spin-1 states equal to the rank of the gauge group. As we will see, one seems to be the most reasonable outcome.

To make this statement more concrete, we will focus on three examples among the one-species theories (see the Appendix for the remaining cases).
For each operator $X_R$, the strength of the binding can be estimated in terms of the following score
\begin{equation}
    \Delta = 2 C_r - C_R\,,
\end{equation}
where the coefficients $C$ are the Casimirs of the related representations.
Channels with $\Delta < 0$ are considered repulsive, hence discarded~\footnote{Intriguingly, for Witten-anomalous theories defined by $\Sp(2N)$ with $r$ equal to the defining representation, the adjoint is the only channel, however with $\Delta = -1/2$. Hence, no condensate can possibly form.}. Among the channels with $\Delta > 0$, the one with the largest score is the MAC. The results of our analysis are summarized in columns 5 to 8 of Table~\ref{tab:pseudoreal_summary}, where we focus only on breaking to regular maximal subgroups of the gauge group \cite{Slansky:1981yr,Feger:2019tvk}.
As concrete examples, we focus on the one-species theory based on $\SU(6)$, and on the theories based on $\Spin(12)$ and $\Spin(11)$, the latter being the exception to the rule of the adjoint. 

The first sample model is based on $\SU(6)$ with $r = {\bf 20}_{A3}$~\cite{Yamaguchi:2018xse,Bolognesi:2019fej,Bolognesi:2021hmg,Yamaoka:2025eej}, so that $R = {\bf 35}_\text{Adj}\ [\Delta = 9/2]$ and ${\bf 175}\ [\Delta = -3/2]$, where ${\bf 175} = (00200)$. The only attractive channel, hence MAC, is $R=\text{Adj}$, which leads to the following breaking patterns:
\begin{itemize}
    \item[A)] $\SU(5)\times \U(1)_X$ under which ${\bf 20} \to {\bf 10}_3 \oplus {\bf \overline{10}}_{-3}$.
    \item[B)] $\SU(4)\times\SU(2)\times\U(1)_X$, under which  ${\bf 20} \to {\bf (4,1)}_{3} \oplus {\bf (\overline{4},1)}_{-3} \oplus {\bf (6,2)}_{0}$.
    \item[C)] $\SU(3)\times\SU(3)\times\U(1)_X$, under which ${\bf 20} \to {\bf (3,\overline{3})}_{-1} \oplus {\bf (\overline{3},3)}_{1} \oplus {\bf (1,1)}_{3} \oplus {\bf (1,1)}_{-3}$.
\end{itemize}
In all three cases, the fermions charged under $\mathcal{H}$ are vector-like and acquiring a mass from the VEV proportional to their $X$-charge, with the exception of case B where the neutral component ${\bf (6,2)}_0$ is pseudoreal and remains massless.
Nevertheless, no condensate can form that breaks $\U(1)_X$, consistently with the fact that the only massless fermion after tumbling is $X$-neutral. In case B, therefore, one more tumbling step occurs when the $\SU(4)$ subgroup confines (having faster running than the $\SU(2)$ factor), leading to a vector-like condensate that would break $\SU(2) \to \U(1)_{X'}$ without massless fermions. Hence, following the path B, one additional massless spin-1 state could emerge.

The second sample model is based on $\Spin(12)$ with the spinorial $r = {\bf 32}_{Spin}$, which leads to $R = {\bf 66}_\text{Adj}\ [\Delta = 13/2]$ and ${\bf 462}\ [\Delta = -3/2]$. Again, the only attractive channel is $R=\text{Adj}$, leading to the following breaking paths:
\begin{itemize}
    \item[A)] $\Spin(10) \times \U(1)_X$, with ${\bf 32} \to {\bf 16}_{-1} \oplus {\bf \overline{16}}_{1}$. \item[B)] $\SU(6) \times \U(1)_X$ with ${\bf 32} \to {\bf 6}_{-2} \oplus {\bf \overline{6}}_{2} \oplus {\bf 20}_0$. 
\end{itemize}
Like for the $\SU(6)$ case, all $X$-charged fermions acquire a mass from the VEV, no condensate can break $\U(1)_X$, and only in path B a massless fermion in a pseudoreal representation remains, which exactly matches the $\SU(6)$ model. Further tumbling would, therefore, lead to additional massless spin-1 states.
    
As a final example, we consider the theory based on $\Spin(11)$ with the spinorial $r = {\bf 32}_{Spin}$. While being similar to the previous one, this case has two distinct attractive channels, as $R = {\bf 11}_F\ [\Delta = 35/4]$, $ {\bf 55}_\text{Adj}\ [\Delta = 19/4]$ and ${\bf 462}\ [\Delta = -5/4]$. The MAC is now given by the defining representation, $R={\bf 11}$, which leads to the path
    \begin{equation}
        \Spin(11) \to \Spin(10) \ \text{with} \ {\bf 32} \to {\bf 16} \oplus {\bf \overline{16}}\,,
    \end{equation}
hence leading to a vector-like theory without massless states. Instead, the Next-to--MAC channel $R=\text{Adj}$ gives:
    \begin{equation}
        \Spin(11) \to \Spin(9) \times \U(1)_X\ \text{with} \ {\bf 32} \to {\bf 16}_{1} \oplus {\bf 16}_{-1}\,,
    \end{equation}
leading to confining vector-like theory plus a massless $\U(1)_X$ gauge boson.
    
All other one-species theories give a simpler result, with the MAC always given by the adjoint and always leading to a massless $\U(1)_X$ gauge boson plus a confining vector-like $\mathcal{H}$, see Appendix. As already mentioned, in principle, one could have more massless spin-1 states, up to full abelianization of the gauge group $\mathcal{G}$.

\subsubsection*{Generalized and discrete symmetries}
\noindent

Higher form (generalized) symmetries have been proposed to be a crucial diagnostic of the phases of gauge theories \cite{Aharony:2013hda,Gaiotto:2014kfa}, as the area or perimeter law of Wilson loops can distinguish between a confining and a Higgsed phase~\cite{Iqbal:2024pee}. For pure gauge theories, the one-form symmetry is the center of the gauge group. Adding matter usually breaks the center symmetry depending on the charge of the representation under the center, i.e. their $N$-ality for $\SU(N)$. In the $\SU(6)$ model, therefore, one has center symmetry $\mathbb{Z}_6$, reduced to a one-form $\mathbb{Z}_3$ by the presence of the ${\bf 20}_{A3}$ representation~\cite{Yamaguchi:2018xse}. For the symplectic group, the center is $\mathbb{Z}_2$ and pseudoreal representations are charged. Therefore, the theories based on $\Sp(2N)$ have trivial one-form symmetry. The same is true for the $\Spin(2N+1)$ theories, while $\Spin(2N)$ ones have a center $\mathbb{Z}_2\times\mathbb{Z}_2$ reduced to a $\mathbb{Z}_2$ one-form by the presence of the spinorial. The analysis is summarized in Table~\ref{tab:discrete_symmetries_summary}, where we see that most theories do not have one-form symmetries, which makes it plausible that the Higgsed and confinement phases are smoothly connected~\cite{Fradkin:1978dv}. 

\begin{table*}[tb]
\centering
\small
\renewcommand{\arraystretch}{1.15}
\begin{tabular}{ll|ccc|cl}
\toprule
\textbf{Gauge group} &
\textbf{Matter} &
\textbf{Center} &
\textbf{Rep. charge} &
\textbf{One-form sym.} &
\textbf{Discrete $\U(1)_A$ remnant} & \textbf{Grav. anomaly}
\\
\midrule
\(\mathrm{Sp}(4)\) &
\(\mathbf{16}_{R11}\) &
\(\mathbb{Z}_2\) &
\(1\) &
trivial &
\(\mathbb{Z}_{12}\) & \( 4\ \text{mod}\ 6 \; : \;\; \mathbb{Z}_{12} \to \mathbb{Z}_4\,, \mathbb{Z}_2\)
\\
\(\mathrm{Sp}(8)\) &
\(\mathbf{48}_{A3}\) &
\(\mathbb{Z}_2\) &
\(1\) &
trivial &
\(\mathbb{Z}_{14}\) & \(6\ \text{mod}\ 7 \; : \;\; \mathbb{Z}_{14} \to \mathbb{Z}_2 \)
\\
\softtheoryrule
\(E_7\) &
\(\mathbf{56}_{F}\) &
\(\mathbb{Z}_2\) &
\(1\) &
trivial &
\(\mathbb{Z}_{12}\) & \( 2\ \text{mod}\ 6 \; : \;\; \mathbb{Z}_{12} \to \mathbb{Z}_4\,, \mathbb{Z}_2\)
\\
\softtheoryrule
\(\mathrm{SU}(6)\) &
\(\mathbf{20}_{A3}\) &
\(\mathbb{Z}_6\) &
\(3\) &
\(\mathbb{Z}_3\) &
\(\mathbb{Z}_6 \to \mathbb{Z}_2\) & \(2\ \text{mod}\ 3 \; : \;\; \mathbb{Z}_6 \to \mathbb{Z}_2\)
\\
\softtheoryrule
\(\mathrm{Spin}(11)\) &
\(\mathbf{32}_{Spin}\) &
\(\mathbb{Z}_2\) &
\(1\) &
trivial &
\(\mathbb{Z}_8\) & \(0\ \text{mod}\ 4\; : \;\; \mathbb{Z}_8\)
\\
\(\mathrm{Spin}(12)\) &
\(\mathbf{32}_{Spin}\) &
\(\mathbb{Z}_2 \times \mathbb{Z}_2\) &
\((1,0)\) &
\(\mathbb{Z}_2\) &
\(\mathbb{Z}_8 \to \mathbb{Z}_{4}\) & \( 0\ \text{mod}\ 4 \; : \;\; \mathbb{Z}_8\)
\\
\(\mathrm{Spin}(13)\) &
\(\mathbf{64}_{Spin}\) &
\(\mathbb{Z}_2\) &
\(1\) &
trivial &
\(\mathbb{Z}_{16}\) & \(0\ \text{mod}\ 8\; : \;\; \mathbb{Z}_{16}\)
\\
\midrule
\(\mathrm{Sp}(6)\) &
\(\mathbf{6}_{F}+\mathbf{14}'_{A3}\) &
\(\mathbb{Z}_2\) &
\(1,1\) &
trivial &
trivial & none
\\
\bottomrule
\end{tabular}
\caption{Discrete symmetries of the pseudoreal gauge theories in Table~\ref{tab:pseudoreal_summary} that are likely outside the conformal window. The one-form symmetry is the subgroup of the center acting trivially on all dynamical fermions. The last two columns give the discrete subgroup of the anomalous $\U(1)_A$ symmetry left unbroken by the instanton and the further breaking induced by the gravitational anomaly~\cite{Csaki:1997aw}. When a non-trivial one-form symmetry is present, this is further broken by a mixed anomaly~\cite{Yamaguchi:2018xse,Bolognesi:2019fej}.  }
\label{tab:discrete_symmetries_summary}
\end{table*}

Chiral discrete symmetries, instead, stem from the anomalous $\U(1)_A$ symmetry, present in all pseudoreal theories. This symmetry is broken by instanton effects to a discrete subgroup~\cite{Csaki:1997aw}, which can be determined by the form of the 't~Hooft operator~\cite{tHooft:1976rip}:
\begin{equation}
    \mathcal{O}_\text{'t Hooft} = \prod_r\ \psi_r^{2T_r}\,.
\end{equation} 
Incidentally, this operator also provides a simple and comprehensive way to understand the Witten anomaly~\cite{Witten:1982fp}, which occurs for pseudoreal $r$ with half-integer Dynkin index: in such cases, the 't~Hooft operator cannot be gauge and Lorentz invariant.

For one-species theories, therefore, we find a chiral discrete symmetry $\mathbb{Z}_{2T_r}$, as listed in the one-to-last column of Table~\ref{tab:discrete_symmetries_summary}. In two cases, based on $\SU(6)$ and $\Spin(12)$, the presence of a non-trivial one-form breaks the remnant symmetry further due to a mixed anomaly~\cite{Yamaguchi:2018xse,Bolognesi:2019fej}. The discrete symmetry also has global anomalies~\cite{Csaki:1997aw}, where the only relevant one involves gravity. The anomaly is given by $(d_r\ \text{mod}\ T_r)$, where $d_r$ is the dimension of $r$, as listed in the second column of Table~\ref{tab:discrete_symmetries_summary}. This type of anomaly must be matched in the IR~\cite{Csaki:1997aw}, \plb{however it has been proven that this cannot be done via a symmetry preserving gapped TQFT due to an obstruction when the anomaly coefficient is non-zero~\cite{Cordova:2019jqi}. Moreover, the absence of fermionic bound states excludes the conventional massless composite matching. Thus, unless a different symmetry preserving gapless phase matches the anomaly, it must be spontaneously broken, which we keep as an assumption throughout the rest of this text.}  The minimal requirement is $\mathbb{Z}_{2T_r} \to \mathbb{Z}_X$, where $X$ is a divisor of $2T_r$ such that 
\begin{equation}
    d_r\ \text{mod}\ X/2 \equiv 0\ \text{mod}\ X/2 
\end{equation} 
or $\text{mod}\ X$ if $X$ is odd, see last column of Table~\ref{tab:discrete_symmetries_summary}. We see that only in the $\Spin$ theories the anomaly vanishes, hence requiring no breaking of the global discrete symmetry.  A corresponding analysis for the likely conformal theories is provided in the Appendix, Table~\ref{tab:discrete_symmetries_summary_app}, for completeness.

For two-species theories, we find that the independent discrete symmetry is $\mathbb{Z}_{\text{gcd} (2T_{r1}, 2T_{r2})}\subset \mathbb{Z}_{\sum_r q_r 2 T_r}$, where gcd indicates the greatest common divisor. As a result, all two-species models do not have a discrete remnant, as they always feature a fermion in the fundamental, which has $2 T_F = 1$.

\subsubsection*{Gauge-invariant operators}

From the tumbling analysis, we found that the MAC is almost always the adjoint, leading to a gapped IR theory with at least one massless spin-1 boson. The only exception is the $\Spin(11)$ model, for which the MAC is in the defining representation, leading to a gapped theory without massless states. The tumbling analysis, however, relies on giving a gauge-breaking VEV to a scalar fermion bilinear, so that the theory is assumed to live in a dynamical Higgsed phase. It would be useful to find a consistent description in terms of gauge-invariant condensates\plb{, assuming that the infrared dynamics can also be described in terms of confinement}. 

Starting with one-species theories, the most straightforward possibility is that the condensate is generated by the four-fermion operator $X_R$ in Eq.~\eqref{eq:XR}, with $R$ in the MAC. Due to the pseudoreal nature of the fermion representation, it is impossible to construct gauge-invariant fermionic operators (the product of an odd number of pseudoreal representation gives pseudoreal representations), hence matching of global anomalies {\`a} la `t~Hooft~\cite{tHooft:1976rip,tHooft:1976snw,tHooft:1979rat} gives a crucial hint: anomalous symmetries must be broken spontaneously. This applies both to the anomaly generated by the one-form symmetry (see fifth column of Table~\ref{tab:discrete_symmetries_summary}) and to the gravitational anomaly (see last column of Table~\ref{tab:discrete_symmetries_summary}). An additional argument based on the strong anomaly has been put forward in~\cite{Bolognesi:2021hmg}, implying that the structure of fermion condensate should be consistent with the `t~Hooft operator $\psi_r^{2T_r}$ acquiring a VEV. \plb{Note that the above arguments are based on the crucial assumptions that the theories confine, hence we do not consider the possibility of symmetry-preserving topological QFTs.}

These arguments have already been thoroughly applied to the $\SU(6)$ theory: the four-fermion interaction $X_R$ can lead to the correct breaking $\mathbb{Z}_6 \to \mathbb{Z}_2$~\cite{Yamaguchi:2018xse}, while the strong anomaly seems to support a bilinear condensate as in the tumbling description~\cite{Bolognesi:2021hmg} due to the fact that the 't~Hooft operator contains six fermions. The $\Sp(8)$ theory is in a similar situation for the breaking $\mathbb{Z}_{14} \to \mathbb{Z}_2$ due to the gravitational anomaly. 
The theories based on $\Sp(4)$ and $E_7$ share the same minimal breaking $\mathbb{Z}_{12} \to \mathbb{Z}_4$, which is compatible with both a VEV for $X_R$ and the strong anomaly argument. The same case occurs for the $\Spin(12)$ theory, where $\mathbb{Z}_8 \to \mathbb{Z}_4$ must occur.
\plb{We should remark that the minimal breaking patterns suggested above are consistent with the residual global symmetry preserved by the tumbling condensate $(\psi_r \psi_r)_R$: the patterns and the gauge-orbit arguments are given in the Appendix.}
Finally, the two theories based on $\Spin(11)$ and $\Spin(13)$ have no anomalies, hence no indications on the form of the fermion condensate can be derived and these theories may be consistent if no fermion condensate forms at all. 

No matter the structure of the fermionic condensate, \plb{if the tumbling hint is correct, we should be able to associate the massless spin-1 state(s) with a gauge-invariant operator that carries no global charges.}
In one-species models, only one two-fermion operator can be constructed
\begin{equation}
    \mathcal{O}_A^\mu = \psi_r^\dagger \sigma^\mu \psi_r\,,
\end{equation}
associated to the current of the anomalous $\U(1)_A$ symmetry. \plb{Is it possible that this operator sources a zero mode, given the gauge anomaly of the associated $\U(1)_A$ current?}
As observed in QCD~\cite{Giacosa:2017ojs}, the axial anomaly gives a large mass to the pseudoscalar $\eta'$ but not to the corresponding spin-1 state $\omega$. The simple reason is that the 't~Hooft operator, which captures the anomaly in the low energy effective theory, carries a non-trivial $\U(1)_A$ charge, as the $\eta'$, while the spin-1 current does not. 
\plb{Hence, we conjecture that $\mathcal{O}_A^\mu$ is the most likely operator to generate at least one of the massless states, and this hypothesis can be tested via numerical methods.}
For completeness, we should mention that other operators can be built via derivatives: operators such as $\psi_r D^\mu \psi_r$ carry $\U(1)_A$ charge and hence will receive a large mass from the anomaly~\footnote{This observation is coherent with the Weinberg-Witten theorem~\cite{Weinberg:1980kq}, even though the $\U(1)_A$ symmetry is broken at quantum level by the gauge anomaly.}; instead operators such as $\psi^\dagger D^2 \sigma^\mu \psi_r$ typically describe higher mass resonances. \plb{Hence, it is reasonable to expect that there is only one operator, $\mathcal{O}_A^\mu$, that can source one zero mode, even though we cannot exclude the presence of multiple zero modes.}


\subsubsection{Two-species $\Sp(6)$ theory}

We now turn our attention to the only two-species theory which is likely to be outside the conformal window. It is based on $\Sp(6)$ with $r_1 = {\bf 6}_F$ and $r_2 = {\bf 14'}_{A3}$. Its anomaly-free global $\U(1)_V$ must be spontaneously broken, as no spin-1/2 composite states can be formed to match the global anomalies \emph{\`a la} 't~Hooft \cite{tHooft:1976rip,tHooft:1976snw,tHooft:1979rat}. The IR spectrum, therefore, must feature a massless Nambu-Goldstone Boson (NGB). Having two representations, there are three possible fermion bilinears, leading to the tumbling channels:
\begin{itemize}
    \item[a)] ${\bf 6} \times {\bf 6}$ with $R = {\bf 21}_\text{Adj}\ [\Delta = -1/2]$;
    \item[b)] ${\bf 14'} \times {\bf 14'}$ with $R = {\bf 21}_\text{Adj}\ [\Delta = 7/2]$ and ${\bf 84}\ [\Delta = -3/2]$;
    \item[c)] ${\bf 6} \times {\bf 14'}$ with $R = {\bf 14} \ [\Delta = 5/2]$ and ${\bf 70} \ [\Delta = -1/2]$.
\end{itemize}
There are only two attractive channels. The MAC is in the adjoint, leading to
\begin{equation}
    \Sp(6) \to \SU(3) \times \U(1)_X\,,
\end{equation}
with
\begin{equation}
    {\bf 6} \to {\bf 3}_{1} \oplus {\bf \overline{3}}_{-1}\,, \quad {\bf 14'} \to {\bf 3}_{-2} \oplus {\bf \overline{3}}_{2} \oplus {\bf 8}_0\,.
\end{equation}
Thus, we obtain a vector-like $\SU(3)$ theory with two massive flavors plus a massless octet (gaugino-like). This theory confines, leaving a massless gauge boson. 

The analysis of the four-fermion operators, however, leads to a richer dynamics due to the presence of the two representations. As the MAC is in the adjoint ${\bf 21}$ of $(\psi_{r_2} \psi_{r_2})_{\bf 21}$, besides $X_R = \langle(\psi_{r_2} \psi_{r_2})_{\bf 21}  (\psi_{r_2} \psi_{r_2})_{\bf 21}\rangle$, one can write a second operator containing the MAC:
\begin{equation}
    X_R' = \langle(\psi_{r_2} \psi_{r_2})_{\bf 21}  (\psi_{r_1} \psi_{r_1})_{\bf 21}\rangle\,.
\end{equation}
A VEV for the MAC, therefore, would induce a VEV for the other adjoint channel $(\psi_{r_1} \psi_{r_1})_{\bf 21}$, even though this channel was not even attractive. These two VEVs, which could be reformulated as condensates for both $X_R$ and $X_R'$, break both $\U(1)_A$ and $\U(1)_V$, hence generating a massless NGB and a massive $\eta'$-like state (together with a massless spin-1 state).

However, the picture drawn above would be inconsistent with the strong anomaly argument, which requires a VEV for the 't~Hooft operator $\langle \psi_{r_1} \psi_{r_2}^5 \rangle$. As this operator can be decomposed as
\begin{equation}
    \langle \psi_{r_1} \psi_{r_2}^5 \rangle \sim \langle (\psi_{r_1} \psi_{r_2})_{\bf 14} (\psi_{r_2} \psi_{r_2})_{\bf 21} (\psi_{r_2} \psi_{r_2})_{\bf 21}  \rangle\,,
\end{equation}
consistency could be restored if we assume that both the MAC and the Next-to-MAC acquire VEVs. This configuration of condensates would, again, break both $\U(1)_A$ and $\U(1)_V$. The breaking of the gauge symmetry allows a maximal unbroken group  (e.g. see~\cite{Cacciapaglia:2024duu}):
\begin{equation}
    \Sp(6) \to \SU(2) \times \U(1) \times \U(1)\,,
\end{equation}
hence predicting the presence of two massless spin-1 states, which could be associated with the two currents $\psi_{r_1}^\dagger \sigma^\mu \psi_{r_1}$ and $\psi_{r_2}^\dagger \sigma^\mu \psi_{r_2}$.

\subsubsection*{Summary and outlook}
\noindent

Pseudoreal theories, defined in terms of an odd number of Weyl fermions in a pseudoreal representation, are poorly understood due to their odd properties. They share common features with the better understood theories: vector-like (real Euclidean path integral) and chiral (absence of non-trivial mass operators) ones. Anomalies~\cite{Witten:1982fp,Wang:2018qoy} allowed to remove many such theories, leaving a list of 15 one-species and 4 two-species minimal models~\cite{Cacciapaglia:2026kry}, see Table~\ref{tab:pseudoreal_summary}.

In this letter, we employ state-of-the-art theoretical machinery to peruse asymptotically free pseudoreal theories and thoroughly analyze their IR dynamics, when the gauge couplings become strong. A number of theories are likely to flow towards an IR fixed point, hence leading to a quantum conformal field theory at low energies. Interestingly, all theories potentially affected by the \WWW anomaly~\cite{Wang:2018qoy} are in this category. For the remaining cases, we employ a tumbling analysis (assuming a dynamical Higgsed phase driven by bilinear fermion operator condensates), matching of the global symmetry anomalies, and analysis of the fermionic gauge-invariant operators. All theories share the following common traits: anomalous global symmetries must be broken due to the absence of fermionic bound states \plb{and of symmetry-preserving gapped TQFT}; the most attractive channel is the adjoint, hence leading to at least one massless spin-1 boson in the deep IR (with one exception); effectively, one can build one such operator in one-species models and two in two-species models, associated to the global $\U(1)$ symmetries.

We identify a consistent picture where one-species theories (except one) lead to confinement with a gapped spectrum, leaving \textit{exactly one massless non-interacting spin-1} bound state in the deep IR. For two-species theories, the most likely scenario predicts \textit{two massless spin-1 states and one Nambu-Goldstone boson}. Discrete anomaly matching, instead, is always consistent with the presence of bilinear condensates.

\vspace{0.5cm}

Our results provide a key step forward in the understanding of quantum field theories, \plb{giving a consistent} picture of the IR dynamics of the poorly understood pseudoreal theories. As such theories are characterized by a real Euclidean path integral, they could be studied on the Lattice, where our predictions can be put to the test. We also foresee the application of the functional renormalization group~\cite{Wetterich:1992yh,Dupuis:2020fhh} to check for the existence of massless spin-1 states. These techniques have recently been applied to chiral gauge theories~\cite{Li:2025tvu,Li:2026ayh}, revealing unexpected dynamical properties.

Finally, some of the theories in Table~\ref{tab:pseudoreal_summary} would retain asymptotic freedom once promoted to $\mathcal{N}=1$ supersymmetry, c.f. third column of the table. Such theories can be analyzed using supersymmetric techniques. For instance, in Ref.~\cite{Csaki:1996zb}, the possible IR dynamics of the supersymmetric $\SU(6)$ theory with $r={\bf 20}_{A3}$ has been derived from a parent S-confining theory: it was found that it features either a runaway superpotential for the supersymmetric analog of $X_R$, or a dynamics without superpotential. The latter may indicate that the IR dynamics is better described in terms of a K\"ahler potential, for instance in terms of a single vector superfield. The analysis of the supersymmetric pseudoreal theories could also provide hints on the dynamics of non-supersymmetric theories via supersymmetry breaking, e.g. see~\cite{Murayama:2021xfj,Csaki:2021xhi}.

\vspace{1cm}

\paragraph{Acknowledgements.}

\noindent
We thank Francesco Sannino for insightful discussions. KK also thanks the LPTHE for hospitality during the initial stages of this project.

\vspace{0.3cm}

\section*{Appendix}

\renewcommand\thetable{A--\Roman{table}}
\setcounter{table}{0}
\renewcommand\theequation{A\arabic{equation}}
\setcounter{equation}{0}

\subsubsection*{Further tumbling analyses}

The complete list of asymptotic free pseudoreal theories is provided in Table~\ref{tab:pseudoreal_summary}. Here we provide details on the tumbling analysis for the theories that are not likely to be inside the conformal window, which are not detailed in the main text. After determining the attractive channels within all possible fermion bilinear operators, we focus only on breakings to regular maximal subgroups of the gauge group.

Among the theories likely to be outside the conformal window, besides the theories based on $\SU(6)$, $\Spin{(12)}$ and $\Spin{(11)}$ presented in the main text, we have:
\begin{itemize}
    \item $E_7$ with $r = {\bf 56}_F$, so that $R = {\bf 133}_\text{Adj}\ [\Delta = 21/2]$ and ${\bf 1463}\ [\Delta = -3/2]$, where ${\bf 1463} = (0000020)$. The only attractive channel is the adjoint, which can only break
    \begin{multline}
        E_7\to E_6 \times \U(1)_X\,, \quad \text{with} \\
        {\bf 56} \to {\bf 27}_{-1} \oplus {\bf \overline{27}}_{1} \oplus {\bf 1}_{3} \oplus {\bf 1}_{-3}\,.
    \end{multline}
    As the condensate is proportional to the generator of $\U(1)_X$, all the above fermions pick up a dynamical mass proportional to their $X$-charge, and the vector-like $E_6$ gauge theory confines without massless states.

    \item $\Spin(13)$ with $r = {\bf 64}_{Spin}$, so that $R={\bf 78}_\text{Adj}\ [\Delta = 17/2]$, ${\bf 286}\ [\Delta = 9/2]$ and ${\bf 1716}\ [\Delta = -3/2]$. The MAC is the adjoint, leading to
    \begin{multline}
        \Spin(13) \to \Spin(11) \times \U(1)_X \quad \text{with} \\ {\bf 64} \to {\bf 32}_{1} \oplus {\bf 32}_{-1}\,.
    \end{multline}
Again, the fermions pick up a mass, leading to a confining vector-like $\Spin(11)$ theory.

The N-MAC $\bf 286$ channel, instead, tumbles to 
\begin{multline}
    \Spin(13) \to \Spin(10) \times \SU(2) \quad \text{with} \\ {\bf 64} \to ({\bf 16}, {\bf 2}) + ({\bf \overline{16}}, {\bf 2})\,,
\end{multline}
with no massless fermions and vector-like confinement.
\end{itemize}

For the two undetermined theories in the symplectic family, we would have the following tumbling patterns:
\begin{itemize}
    \item $\Sp(4)$ with ${r = \bf 16}_{R11}$, so that $R={\bf 10}_\text{Adj}\ [\Delta = 9/2]$, ${\bf 35}\ [\Delta = 3/2]$ and ${\bf 81}\ [\Delta = -5/2]$, where ${\bf 35} = (2,1)$ and ${\bf 81} = (2,2)$. The MAC is the adjoint, leading to
    \begin{multline}
        \Sp(4) \to \SU(2)\times \U(1)_X \quad \text{with} \\ {\bf 16} \to {\bf 4}_{1} \oplus {\bf 4}_{-1} \oplus {\bf 2}_{3} \oplus {\bf 2}_{-3} \oplus {\bf 2}_{1} \oplus {\bf 2}_{-1}\,.
    \end{multline}
    All fermions charged under $\U(1)_X$ acquire a mass, and the vector-like $\SU(2)$ confines.

    The N-MAC $\bf 35$ channel, instead, would tumble to a non-maximal subgroup of $\Sp(4)$.
    
    \item $\Sp(8)$ with $r = {\bf 48}_{A3}$, so that $R={\bf 36}_\text{Adj}\ [\Delta = 23/2]$, ${\bf 315}\ [\Delta = 5/2]$ and ${\bf 825}\ [\Delta = -3/2]$. The MAC in the adjoint channel leads to
    \begin{multline}
        \Sp(8) \to \SU(4) \times \U(1)_X\quad \text{with}\\ {\bf 48} \to {\bf 20}_{1} \oplus {\bf \overline{20}}_{-1} \oplus {\bf 4}_{-3} \oplus {\bf \overline{4}}_{3}\,,
    \end{multline}
    once again leading to a vector-like confining $\SU(4)$ theory.

    The N-MAC $\bf 315$ channel would necessarily tumble to a non-maximal subgroup of $\Sp(8)$.
\end{itemize}

\vspace{1cm}

\begin{table*}[tb]
\centering
\small
\renewcommand{\arraystretch}{1.15}
\begin{tabular}{ll|ccc|cl}
\toprule
\textbf{Gauge group} &
\textbf{Matter} &
\textbf{Center} &
\textbf{Rep. charge} &
\textbf{One-form sym.} &
\textbf{Discrete $\U(1)_A$ remnant} & \textbf{Grav. anomaly}
\\
\midrule
\(\mathrm{SU}(2)\) &
\(\mathbf{4}_{S3}\) (*) &
\(\mathbb{Z}_2\) &
\(1\) &
trivial &
\(\mathbb{Z}_{10}\) & \( 4\ \text{mod}\ 5 \; : \;\; \mathbb{Z}_{10} \to \mathbb{Z}_2\)
\\
\(\mathrm{Sp}(6)\) &
\(\mathbf{56}_{S3}\) (*) &
\(\mathbb{Z}_2\) &
\(1\) &
trivial &
\(\mathbb{Z}_{36}\) & \(2\ \text{mod}\ 18 \; : \;\; \mathbb{Z}_{36} \to \mathbb{Z}_4\,, \mathbb{Z}_2 \)
\\
\(\mathrm{Sp}(6)\) &
\(\mathbf{64}_{R11}\) &
\(\mathbb{Z}_2\) &
\(1\) &
trivial &
\(\mathbb{Z}_{32}\) & \(0\ \text{mod}\ 16 \; : \;\; \mathbb{Z}_{32}\)
\\
\(\mathrm{Sp}(10)\) &
\(\mathbf{132}_{A5}\) &
\(\mathbb{Z}_2\) &
\(1\) &
trivial &
\(\mathbb{Z}_{42}\) & \(6\ \text{mod}\ 21 \; : \;\; \mathbb{Z}_{42} \to \mathbb{Z}_6\,, \mathbb{Z}_2 \)
\\
\(\mathrm{Sp}(12)\) &
\(\mathbf{208}_{A3}\) &
\(\mathbb{Z}_2\) &
\(1\) &
trivial &
\(\mathbb{Z}_{44}\) & \(10\ \text{mod}\ 22 \; : \;\; \mathbb{Z}_{44} \to \mathbb{Z}_4\,, \mathbb{Z}_2 \)
\\
\(\mathrm{Sp}(16)\) &
\(\mathbf{544}_{A3}\) &
\(\mathbb{Z}_2\) &
\(1\) &
trivial &
\(\mathbb{Z}_{90}\) & \(4\ \text{mod}\ 45 \; : \;\; \mathbb{Z}_{90} \to  \mathbb{Z}_2 \)
\\
\softtheoryrule
\(\mathrm{Spin}(19)\) &
\(\mathbf{512}_{Spin}\) &
\(\mathbb{Z}_2\) &
\(1\) &
trivial &
\(\mathbb{Z}_{128}\) & \(0\ \text{mod}\ 64\; : \;\; \mathbb{Z}_{128}\)
\\
\(\mathrm{Spin}(20)\) &
\(\mathbf{512}_{Spin}\) &
\(\mathbb{Z}_2 \times \mathbb{Z}_2\) &
\((1,0)\) &
\(\mathbb{Z}_2\) &
\(\mathbb{Z}_{128} \to \mathbb{Z}_{64}\) & \( 0\ \text{mod}\ 64 \; : \;\; \mathbb{Z}_{128}\)
\\
\bottomrule
\end{tabular}
\caption{Discrete symmetries of the one-species pseudoreal gauge theories in Table~\ref{tab:pseudoreal_summary} that are likely inside the conformal window. We follow the same scheme as in Table~\ref{tab:discrete_symmetries_summary}. For the two-species theories, no non-trivial discrete symmetries apply.}
\label{tab:discrete_symmetries_summary_app}
\end{table*}

The one-species theories that are likely to be inside the conformal window also have non-trivial discrete symmetries, which we list in Table~\ref{tab:discrete_symmetries_summary_app}. We recall that the remaining two-species theories have trivial discrete symmetries.

\subsubsection*{Perturbative fixed points}
\label{app:perturbative}

\plb{As a check of the conformal-window estimates, we compute the
two- and three-loop gauge beta functions for the theories in
Table~\ref{tab:pseudoreal_summary}. With \(a=g^2/(16\pi^2)\), we write
\begin{equation}
\beta(g)=-\frac{g^3}{16\pi^2}
\left(\beta_0+\beta_1 a+\beta_2 a^2\right),
\end{equation}
where \(\beta_0\) is given in Eq.~\eqref{eq:b}. For Weyl fermions,
\begin{align}
\beta_1={}&\frac{34}{3}C_G^2
 -\sum_r n_rT_r
 \left(\frac{10}{3}C_G+2C_r\right),\\
\beta_2={}&\frac{2857}{54}C_G^3
 -\frac{1415}{54}C_G^2T
 +\frac{79}{54}C_GT^2 \notag\\
 &-\frac{205}{18}C_GC
 +\frac{11}{9}TC+D .
\end{align}
Here \(T=\sum_r n_rT_r\), \(C=\sum_r n_rT_rC_r\), and
\(D=\sum_r n_rT_rC_r^2\). The three-loop coefficient is given
in the \(\overline{\mathrm{MS}}\) scheme~\cite{Pickering:2001aq}.
Table~\ref{tab:perturbative} lists the positive infrared-attractive
zeros at two and three loops, expressed as
\(\alpha_*^{(\ell)}=g_*^2/(4\pi)\). A dash indicates that no such
zero exists at that order.}

\begin{table}[tb]
\centering
\footnotesize
\setlength{\tabcolsep}{2pt}
\begin{tabular}{@{}llccc@{}}
\toprule
\textbf{Gauge group} &
\textbf{Matter} &
\textbf{CW} &
$\alpha_*^{(2)}$ &
$\alpha_*^{(3)}$ \\
\midrule
$\SU(2)$ &
$\mathbf{4}_{S3}$ (*) &
likely & 1.971 & 1.047 \\

$\Sp(4)$ &
$\mathbf{16}_{R11}$ &
? & 29.322 & 1.729 \\

$\Sp(6)$ &
$\mathbf{56}_{S3}$ (*) &
likely & 0.111 & 0.100 \\

$\Sp(6)$ &
$\mathbf{64}_{R11}$ &
likely & 0.251 & 0.197 \\

$\Sp(8)$ &
$\mathbf{48}_{A3}$ &
? & -- & -- \\

$\Sp(10)$ &
$\mathbf{132}_{A5}$ &
likely & 0.265 & 0.187 \\

$\Sp(12)$ &
$\mathbf{208}_{A3}$ &
likely & 0.431 & 0.241 \\

$\Sp(16)$ &
$\mathbf{544}_{A3}$ &
likely & 0.026 & 0.024 \\

\midrule
$E_7$ &
$\mathbf{56}_{F}$ &
out & -- & -- \\

$\SU(6)$ &
$\mathbf{20}_{A3}$ &
out & -- & -- \\

$\Spin(11)$ &
$\mathbf{32}_{Spin}$ &
out & -- & -- \\

$\Spin(12)$ &
$\mathbf{32}_{Spin}$ &
out & -- & -- \\

$\Spin(13)$ &
$\mathbf{64}_{Spin}$ &
out & -- & -- \\

\midrule
$\Spin(19)$ &
$\mathbf{512}_{Spin}$ &
likely & 0.080 & 0.058 \\

$\Spin(20)$ &
$\mathbf{512}_{Spin}$ &
likely & 0.091 & 0.063 \\

\midrule
$\Sp(4)$ &
$\mathbf{4}_{F}+\mathbf{20}_{S3}$ (*) &
likely & 0.386 & 0.297 \\

$\Sp(6)$ &
$\mathbf{6}_{F}+\mathbf{14}'_{A3}$ &
out & -- & -- \\

$\Sp(10)$ &
$\mathbf{10}_{F}+\mathbf{110}_{A3}$ &
likely & 2.793 & 0.612 \\

$\Sp(14)$ &
$\mathbf{14}_{F}+\mathbf{350}_{A3}$ &
likely & 0.116 & 0.092 \\
\bottomrule
\end{tabular}
\caption{Positive infrared-attractive zeros of the two- and
three-loop beta functions for the theories of
Table~\ref{tab:pseudoreal_summary}. The CW column reproduces
the conformal-window estimates of that table. A dash indicates
that no such zero exists at that order.}
\label{tab:perturbative}
\end{table}

\subsubsection*{Discrete symmetry of tumbling configurations}

\plb{We now calculate the residual discrete symmetry preserved by the tumbling condensate $B_R=(\psi_r\psi_r)_R$. Denote the instanton remnant as $\mathbb{Z}_{N}$ with $N=2T_r$ acting on the fermion as $\psi_r\to\omega\psi_r$, where $\omega=e^{2\pi i/N}$. The bilinear condensate then transforms as $B_R\to \omega^2 B_R$. A global discrete transformation is preserved when the transformed condensate lies in the same gauge orbit as the original one. In particular if $4$ divides $N$, the transformation $\psi_r\to i
\psi_r$ sends $B_R\to -B_R$ and remains unbroken if $-B_R$ is in the same gauge orbit as $B_R$.
For an adjoint bilinear, this is satisfied when $-1$ belongs to the Weyl group of the gauge group. This holds for the relevant $\Sp$, $E_7$, and $\Spin$
theories. In the symplectic case, the sign reversal can be
seen explicitly. Writing a Cartan element as
$H=\operatorname{diag}(D,-D)$, the matrix representing the
symplectic form in the defining representation obeys
\begin{equation}
J=
\begin{pmatrix}
0&I_n\\
-I_n&0
\end{pmatrix},
\qquad JHJ^{-1}=-H .
\end{equation}
The vector channel of $\Spin (11)$ also admits this sign reversal. This understood easily by noting that any vector can be rotated to minus itself. 
The resulting patterns are summarized in
Table~\ref{tab:tumbling_discrete} and are also reproduced by a four-fermion condensate $X_R$.
\begin{table}[tb]
\centering
\small
\setlength{\tabcolsep}{3pt}
\begin{tabular}{@{}llcc@{}}
\toprule
\textbf{Gauge group} &
\textbf{Matter} &
\textbf{Instanton remnant} &
\textbf{After tumbling} \\
\midrule
$\Sp(4)$ &
$\mathbf{16}_{R11}$ &
$\mathbb{Z}_{12}$ &
$\mathbb{Z}_4$ \\
$\Sp(8)$ &
$\mathbf{48}_{A3}$ &
$\mathbb{Z}_{14}$ &
$\mathbb{Z}_2$ \\
$E_7$ &
$\mathbf{56}_{F}$ &
$\mathbb{Z}_{12}$ &
$\mathbb{Z}_4$ \\
$\SU(6)$ &
$\mathbf{20}_{A3}$ &
$\mathbb{Z}_{6}$ &
$\mathbb{Z}_2$ \\
$\Spin(11)$ &
$\mathbf{32}_{Spin}$ &
$\mathbb{Z}_{8}$ &
$\mathbb{Z}_4$ (adjoint) \\
$\Spin(11)$ &
$\mathbf{32}_{Spin}$ &
$\mathbb{Z}_{8}$ &
$\mathbb{Z}_4$ (vector) \\
$\Spin(12)$ &
$\mathbf{32}_{Spin}$ &
$\mathbb{Z}_{8}$ &
$\mathbb{Z}_4$ \\
$\Spin(13)$ &
$\mathbf{64}_{Spin}$ &
$\mathbb{Z}_{16}$ &
$\mathbb{Z}_4$ \\
\bottomrule
\end{tabular}
\caption{Discrete chiral subgroups preserved by the putative
tumbling configurations.}
\label{tab:tumbling_discrete}
\end{table}}

\end{document}